\documentstyle[latexsym,epsf,amsfonts,prd,aps,twocolumn]{revtex}

\newcommand{\be}[1]{\begin{equation}\label{#1}}
\newcommand{\ee}{\end{equation}}
\newcommand{\ba}[1]{\begin{eqnarray}\label{#1}}
\newcommand{\ea}{\end{eqnarray}}
 
\newcommand{\ket}[1]{\left|#1\right>} 
\newcommand{\bra}[1]{\left<#1\right|}

\begin{document} 
 
\title{Addendum to: 
"On the anomalous dimension for the transversity distribution"}  
\author{Ralf Kuhn$^{1,2}$, Michael Meyer-Hermann$^{1,*}$, and Ralf Sch\"utzhold$^1$}
\address{$^1$Institut f\"ur Theoretische Physik, Technische  Universit\"at
Dresden, D-01062  Dresden, Germany\\
$^2$Max-Planck-Institut f\"ur Physik komplexer Systeme, 
D-01187 Dresden, Germany\\
$^*$ Electronic address: meyer-hermann@physik.tu-dresden.de}
\date{\today}
\maketitle

\begin{abstract} 
We review the results of our previous publication 
[Phys.\ Rev.\ D {\bf 63}, 116001 (2001)] in the light
of recent calculations and comments. 
\end{abstract} 
   

\bigskip

In our previous calculation \cite{Mey01} we applied the method of
Ioffe and Khodjamirian \cite{Iof95} to calculate the leading order
anomalous dimension of the twist-2 transversity distribution.
The original motivation was to get an independent derivation of the
splitting function at Bjorken-$x=1$ -- which has not been 
calculated directly (as no corresponding sum rule exists) but 
fixed by physical arguments in \cite{Art90}. 
The obtained result differed from the one found in literature.
This discrepancy offered two alternative implications: Either
the anomalous dimension deviates from the expected result at $x=1$
or the used method \cite{Iof95} is not applicable in this way.

Recent calculations \cite{Muk01,private} (see also \cite{Rat01} for a
review), which are 
based on techniques different from the one we used, e.g.~the
light-front Hamiltonian approach, also derive the behavior at $x=1$
directly. In this way the original result for the anomalous dimension
\cite{Art90} has been confirmed.

From this perspective, one is lead to review the method \cite{Iof95}
in order to find
the reason for the above mentioned discrepancy.
In reference \cite{Iof95} it was derived that the matrix element
\ba{defoperh1}
\int \frac{d\lambda}{4\pi}\,e^{i\lambda x}
\bra{p,s} \overline{\psi}(0) \sigma_{\mu\nu} i\gamma_5 \psi(\lambda n)
\ket{p,s}
\,,
\ea
which defines the transversity distribution, is {\it directly}
related to the forward scattering amplitude
\ba{h1Tmu}
\frac{i}{2}
\int d^4y\,e^{iqy}
\bra{p,s} {\cal T}\left[j_{\mu 5}(y) j(0)
 + j(y) j_{\mu 5}(0) \right]
\ket{p,s}
\ea
via Eq.\,(11) in \cite{Iof95}. The relation is restricted to
leading order in the QCD-coupling and its derivation is based
on the assumption of light cone dominance, see Eq.\,(5) in
\cite{Iof95}. Using the definition of the transversity distribution
Eq.\,(\ref{defoperh1}) the divergent part, i.e.~the corresponding 
splitting function, follows directly (see Eq.~(20) in \cite{Muk01}).
If the identity derived in \cite{Iof95} holds strictly in the
required order, the divergent part of the forward scattering
amplitude Eq.~(\ref{h1Tmu}) should be same. We calculated
this divergent part and found a deviation \cite{Mey01}.

As one possible interpretation one may suspect that the divergent part
is changed due to the assumption of light cone dominance. Indeed, the 
singularities of the exact quark propagator differ from the
free two-point function used in \cite{Iof95}. This could
imply, that the identity Eq.\,(11) in \cite{Iof95} is
not applicable to the divergent parts. 
Nevertheless,
recently it was argued that one can recover the correct
divergent part by improving the original method through
renormalization group arguments 
(for details see \cite{Blu01}).


The leading order splitting function of the transversity distribution 
is of special relevance not only for the calculation of the
next-to-leading order splitting function but also for experiments
aiming at measurements of $h_1$ which are currently in preparation
(e.g.~at HERMES, see \cite{Jaf97}). 
In order to avoid misunderstandings we would like to emphasize
that the outcome of our previous article \cite{Mey01} pointed
primarily toward conceptual problems.
Especially in view of recent independent calculations
there is no reason to question the original result for 
the splitting function.

{\bf Acknowledgment} 

We thank P.~G.~Ratcliffe, G.~Soff, and J.~Bl\"umlein for helpful
comments and criticism.
This work was partially supported by BMBF, DFG, and GSI.
R.~S.~acknowledges financial support by the Alexander-von-Humboldt
foundation.  


\end{document}